\documentclass[12pt]{iopart}
\pdfoutput=1
\usepackage{dcolumn}
\usepackage{bm}
\usepackage{graphicx}
\usepackage{color}
\usepackage{subfigure}
\usepackage{hyperref}
\usepackage{latexsym}
\usepackage{amsthm}
\usepackage{amssymb}
\usepackage{cite}
\DeclareGraphicsExtensions{.jpg,.pdf, .mps, .png, .eps, .ps, .EPS,.gif}

\begin{document}
\def\be{\begin{equation}}
\def\ee{\end{equation}}

\def\bc{\begin{center}}
\def\ec{\end{center}}
\def\bea{\begin{eqnarray}}
\def\eea{\end{eqnarray}}
\newcommand{\avg}[1]{\langle{#1}\rangle}
\newcommand{\Avg}[1]{\left\langle{#1}\right\rangle}

\def\ie{\textit{i.e.}}
\def\etal{\textit{et al.}}
\def\m{\vec{m}}
\def\G{\mathcal{G}}

\title[The spectral dimension of   simplicial complexes]{ The spectral dimension of  simplicial complexes:\\ \ \  a  renormalization group theory}

\author{Ginestra Bianconi}

\address{
School of Mathematical Sciences, Queen Mary University of London, London, United Kingdom\\Alan Turing Institute, The British Library, London, United Kingdom\\}
\ead{g.bianconi@qmul.ac.uk}

\author{Sergey N. Dorogovstev}

\address{Departamento  de  F\'isica  da  Universidade  de  Aveiro \&I3N,Campus  Universit\'ario  de  Santiago,  3810-193  Aveiro,  Portugal\\A.F.  Ioffe  Physico-Technical  Institute,  194021  St.   Petersburg,  Russia}
\vspace{10pt}
\begin{indented}
\item[]
\end{indented}

\begin{abstract}
Simplicial complexes are increasingly used to study complex system structure and dynamics including diffusion, synchronization and epidemic spreading.
The spectral dimension of the graph Laplacian is known to determine the  diffusion properties at long time scales. Using   the renormalization group here we calculate the  spectral dimension of the graph Laplacian of two classes of non-amenable $d$ dimensional simplicial complexes: the Apollonian networks and the pseudo-fractal networks. We analyse the scaling of the spectral dimension with the topological dimension $d$ for $d\to \infty$ and we point out that randomness such as the one present in Network Geometry with Flavor can diminish the value of the spectral dimension of these structures.
\end{abstract}

\section{Introduction}

Simplicial complexes \cite{Perspective,Lambiotte1,Bassett,top1} are generalized network structures that are ideal to investigate network geometry and topology. They are formed by simplices like nodes, links, triangles, tetrahedra etc. that describe the higher order interactions between the elements of a complex system. The underlying decomposition of simplicial complexes in their geometric building blocks (the simplices) allows to answer novel questions in network topology and geometry. Network geometry and topology are  emergent topics in statistical mechanics and applied mathematics that  explore the properties of complex interacting systems using geometrical concepts and methods tailored to the discrete setting. Novel equilibrium and non-equilibrium modelling frameworks for simplicial complexes have been proposed recently in the literature \cite{Emergent,Hyperbolic,Flavor,Polytopes,Equilibrium,Triangulations,Doro_link,Petri}, and some classic example of deterministic networks such as Apollonian networks \cite{apollonian,apollonian2} and pseudo-fractal networks\cite{pseudo_Doro}  can be reinterpreted as skeleton of simplicial complexes. Additionally we note that simplicial complexes have been already extensively used in the quantum gravity literature to describe quantum space-time. 
They constitute, for instance, the underlying structures of Causal Dynamical Triangulations and of Tensor networks   \cite{Loll,Rivasseau}.
From the applied point of view tools of network topology and geometry span brain research \cite{Bassett,Vaccarino}, financial networks \cite{Aste1}, social science \cite{Petri} and condensed matter\cite{Nanoparticles}.

Network geometry and topology have been recently shown to have a very significant effect on dynamics including synchronization dynamics\cite{Ana,Ana2,Arenas}, epidemic spreading \cite{Iacopini,Kahng_SIS,Arenas2} and percolation \cite{BianconiZiff,BianconiZiff2,BianconiZiff3}.
The spectral properties of network geometries  constitute a direct link to the diffusion processes defined on the same structures. The spectral properties  of networks have been extensively studied in the literature \cite{Samukhin1,Samukhin2,Doro_book,Erzan,Ising,Enzo_Lucas}, however the study of the spectral properties of discrete network geometries is only at its infancy.
Here we focus on the  spectral dimension \cite{Toulouse,Kahng,Kim,Burioni1,Burioni2,Burioni3} of the network geometry, which is known to determine the return distribution of a random walk and define universality classes of the Gaussian model. Recently the spectral dimension has been shown to be key to characterize the stability of the synchronized phase in the Kuramoto model defined on simplicial complexes \cite{Ana,Ana2}.
The spectral dimension \cite{Toulouse} is a concept that extends the notion of dimension for a lattice and in fact it is equal to the lattice dimension for Euclidean lattices.
The  spectral dimension $d_S$ is however distinct from the Hausdorff dimension $d_H$ for a general network \cite{Jonsson1,Jonsson2}.
This concept has been introduced to study the diffusion on fractal structures \cite{Toulouse} and it has been then applied to a variety of contexts including the characterization of the stability of the 3D folding of proteins \cite{Burioni3}.
Interestingly, the notion of spectral dimension is used  widely in  quantum gravity to compare different models of quantum space-time in search for their universal properties\cite{Loll2,dario1,dario2}. 

In this work we  investigate the spectral properties of the non-amenable skeleton of $d$-dimensional  simplicial complexes generated by  deterministic and random models: the Apollonian \cite{apollonian}, the pseudo-fractal \cite{pseudo_Doro} simplicial complexes and the Network Geometry with Flavor (NGF) \cite{Flavor,Hyperbolic,Polytopes}. The Apollonian simplicial complexes and the NGF with flavor $s=-1$ are hyperbolic manifolds, while the other studied simplicial complexes have a non-amenable hierarchical structure. These  discrete network structures are ideal to perform real-space renormalization group calculations revealing the critical properties of percolation \cite{BianconiZiff,BianconiZiff2,BianconiZiff3,Patchy,BoettcherZiff,percolation_apollonain,Havlin1,Havlin2} the Ising model \cite{Boettcher_RG} and Gaussian model \cite{Burioni2,Kahng,Kim}. Here we use the renormalization group technique proposed in Ref. \cite{Kahng,Kim} to predict the spectral dimension of simplicial complexes of different topological dimension $d$ for the Apollonian and the pseudo-fractal network. Moreover we will compare numerically the spectral properties of these deterministic network models with the spectral properties of the simplicial complexes generated by the model Network Geometry with Flavor \cite{Flavor} which include some relevant randomness. 
Our results reveal that the spectral dimension of the deterministic networks can be higher than the topological dimension. Specifically we see that planar Apollonian networks (in $d=3$)  have a  spectral dimension  $d_S=3.73..$..
Additionally we found that  the spectral dimension $d_S$ grows asymptotically for large $d$ as $d_S\simeq 2d\ln d$ for both the Apollonian and the pseudo-fractal network. 
  Finally we show numerically that topological randomness can diminish significantly the spectral dimension of the networks.

  The paper is organized as follows:
  In Sec. 2 we define simplicial complexes and  introduce the simplicial complex models investigated in this work. In Sec. 3 we define the spectral dimension and the relation between the spectral dimension and the Gaussian model. In Sec. 4 we introduce the real space renormalization group approach used in this work. In Sec. 5 we derive the RG equations for the Gaussian model on the Apollonian network,  we theoretically predict the spectral dimension of Apollonian networks of any dimension $d\geq 2$ and we compare the spectral properties of Apollonian networks with the spectral properties of NGFs with flavor $s=-1$. In Sec. 6 we use the RG approach to predict the spectral dimension of pseudo-fractal networks of any topological dimension $d\geq 2$ and we compare the results with numerical result of both pseudo-fractal networks and NGFs with flavor $s=0$ and $s=1$. Finally in Sec. 7 we provide the conclusions.

\section{Simplicial complexes under study}

\subsection{General remarks}

\subsection{Simplicial complexes}

A simplicial complex of $N$ nodes can be used to   describe complex interacting systems including higher order interactions. A simplicial complex is formed by simplices glued along their faces.
A $\delta$-dimensional simplex is a set of $\delta+1$ nodes characterizing a single many body interaction.   
A $0$-simplex is a node, a $1$-simplex is a link, a $2$-simplex is a triangle, and so on.
For instance a $2$-simplex in a collaboration network can indicate that three authors have co-authored a paper, or a $3$-simplex in a face-to-face interaction indicates a group of four people in a conversation.
A $\delta^{\prime}$-dimensional {\em face} $\alpha^{\prime}$ of a  $\delta$-simplex $\alpha$,  is a simplex formed by a subset of $\delta^{\prime}+1$ nodes of $\alpha$, i.e. $\alpha^{\prime}\subset \alpha$.

A $d$-dimensional {\em simplicial complex} ${\mathcal K}$  is formed by a set of simplices of dimensions $0\leq \delta\leq d$ (including at least a $d$-dimensional simplex) that satisfy the two conditions:
\begin{itemize}
\item[(a)] if a simplex $\alpha$ belongs to the simplicial complex, i.e. $\alpha\in {\mathcal K}$ then also all its faces $\alpha^{\prime}\subset \alpha$ belong to the simplicial complex, i.e. $\alpha^{\prime}\in {\mathcal K}$;
\item[(b)] if two simplices $\alpha$ and $\alpha^{\prime}$ belong to the simplicial complex, i.e. $\alpha\in {\mathcal K}$ and $\alpha^{\prime}\in {\mathcal K}$, then either the two simplices do not intercept $\alpha\cap\alpha^{\prime}=\emptyset$
or their intersection is a face of the simplicial complex, i.e. $\alpha\cap\alpha^{\prime}\in{\mathcal K}$.
\end{itemize}

A {\em pure} $d$-dimensional simplicial complex is  only formed by $d$-dimensional simplices and their faces.

 The {\em $1$-skeleton} of a simplicial complex is the network formed exclusively by the nodes and the links or the simplicial complex.

Here we will focus exclusively on  the skeleton  of pure $d$-dimensional simplicial complexes. The simplicial complexes  that we will consider  are the Apollonian,  the pseudo-fractal simplicial complexes and the Network Geometry with Flavor.
In the following paragraphs we will introduce each one of these models.

\subsection{Apollonian network}

A $d$-dimensional Apollonian network \cite{apollonian,apollonian2} (with $d\geq 2$) is the skeleton of a simplicial complex that is generated iteratively by starting  from a single $d$-simplex at iteration $n=0$ and at each  iterations $n>0$ adding a $d$-simplex to every $(d-1)$-dimensional face introduced at the previous generation.
Therefore in these Apollonian networks, at generation $n>0$ there are $N_n$ nodes,  and ${\mathcal N}_n$ nodes added at iteration $n$ with 
\bea
N_n&=&(d+1)\frac{d^n+d-2}{d-1},\nonumber\\
{\mathcal N}_n&=&(d+1)d^{n-1}.
\label{NA}
\eea
The Apollonian networks are small-world, i.e. their Hausdorff dimension is infinity,
\bea
d_H=\infty,
\eea
therefore at each iteration their diameter  grows logarithmically with the total number of nodes of the network.
Moreover, the Apollonian networks of dimension $d$ are manifolds that define discrete hyperbolic lattices.

Let us add here a pair of additional combinatorial properties of Apollonian networks that will be useful in the future.
At each iteration $n$ we call {\em links  of type} $\ell$ the links added at generation $m=n-\ell$.
At generation $n$, the number of $d$-simplices of  generation $n$ attached to links  of type $\ell$ is given by
\bea
w_{\ell}=(d-1)(d-2)^{\ell-1}.
\label{wl}
\eea
The number of $d$-dimensional simplices of generation $n$ incident to nodes    added  at generation $m=n-\ell$ is given by 
\bea
v_{\ell}=d(d-1)^{\ell-1}
\eea
for $\ell>0$.
Moreover it can be easily shown that the  number of links of generation $n$ incident to   nodes added  at generation $m=n-\ell$ is given by 
$v_{\ell}$ for $\ell>0$ and  $v_0=d$ for $\ell=0$.

\subsection{Pseudo-fractal network of any dimension}
The pseudo-fractal network \cite{pseudo_Doro} is the skeleton of a simplicial complex constructed iteratively starting at iteration $n=0$ from a single $d$-simplex (here and in the following we take $d\geq 2$)
At each time $n>0$ we attach a $d$-simplex to every $(d-1)$-dimensional face introduced a time $n\geq 0$.
At iteration $n>0$ the  number of nodes ${N}_n$ and the 
number of links $L_n^{(1)}$ added at iteration $n$ is 
\bea
{N}_n=d+\frac{1}{d}\left[(d+1)^{n}-1\right],\nonumber \\
{\mathcal N}_n=(d+1)^{n}
\label{NP}
\eea
The pseudo-fractal networks are small-world, i.e. their Hausdorff dimension is infinity,
\bea
d_H=\infty.
\eea
As for the Apollonian network, also for the pseudo-fractal networks, at each iteration $n$ we call {\em link of type} $\ell$ the links added at generation $m=n-\ell$.
We observe that at generation $n$ the number of $d$-simplices of   generation $n$ attached to links  of type $\ell$ is given by
\bea
\hat{w}_{\ell}=(d-1)\sum_{\ell^{\prime}=0}^{\ell}(d-2)^{\ell^{\prime}-1}.
\eea
It is also possible to show with straightforward combinatorial arguments that  the  number of $d$-dimensional simplices added to nodes of generation $m=n-\ell$ is 
\bea
\hat{v}_{\ell}=d\sum_{\ell^{\prime}=0}^{\ell}(d-1)^{\ell^{\prime}-1}.
\eea
for $\ell>0$.
Finally the number of links of iteration $n$ added to nodes of generation $m=n-\ell$ is given by $\hat{v}_{\ell}$ for $\ell>0$ and $\hat{v}_0=d$ for $\ell=0$.

\subsection{Network Geometry with Flavor}
Network Geometry with Flavor (NGF) \cite{Flavor,Hyperbolic} generates growing $d$-dimensional simplicial complexes as Apollonian  and pseudo-fractal simplicial complexes. However the dynamics of the NGF is not deterministic but random. 

We assign to every $(d-1)$-dimensional face of a simplex an  {\em incidence number} $\hat{n}_{\alpha}$ equal to the number of $d$-dimensional simplices incident to it minus one. Therefore we note that the incidence number can change with time.

The evolution of the  NGF is dictated by a parameter called  {\em flavor} $s\in \{-1,0,1\}$ 
The algorithm that determines the NGF evolution assumes that 
 at time $t = 1$ the simplicial complex is formed by a  single $d$-simplex. At each time $t > 1$  a $(d-1)$-face $\alpha$ is chosen with probability 
\begin{equation}
\Pi_{d,d-1}(\alpha)=\frac{(1+sn_\alpha)}{Z^{[s]}},
\label{Pa}
\end{equation}
where $Z^{[s]}$ is called the {\em partition function} of the NGF and is given by 
\bea
Z^{[s]}(t)=\sum_{\alpha^{\prime}\in S_{d,d-1}} 
(1+sn_{\alpha^{\prime}}).
\eea
We note here that the Hausdorff dimension of NGFs defined above is always 
\bea
d_{H}=\infty,
\eea
as these networks are small-world for any value of the flavor $s\in\{-1,0,1\}$.
In the case $s=-1$ the NGF  evolves as a subgraph of the Apollonian network connected to the initial $d$-dimensional simplex. In this case  we obtain a random Apollonian network \cite{apollonian2}. Therefore it is interesting to compare the spectral properties of the NGF with $s=-1$  to the spectral properties of the Apollonian network. The NGF with flavor $s=-1$ describe emergent hyperbolic geometries \cite{Hyperbolic}. In fact they are hyperbolic networks emerging from a fully stochastic dynamics that makes no reference to their underlying geometry. Indeed NGF with flavor $s=-1$ form a subset of the Apollonian networks of the same dimension $d$.
In the case $s=0$ and $s=1$ every $(d-1)$-dimensional face can be incident to an arbitrary number of $d$-dimensional simplices. Therefore it is interesting to compare the spectral properties of the NGF with $s=0$ and $s=1$ to the spectral properties of the pseudo-fractal network.
Note that the Network Geometry with Flavor \cite{Flavor} was originally defined with an additional dependence to  another parameter called {\it inverse temperature} $\beta$. Here  we focus only on the case $\beta=0$, therefore we do not need to introduce this additional parameter in this work.

\section{Laplacian spectrum and the Gaussian model}

\subsection{Spectral dimension}

For a network it is possible to defined both a normalized and a un-normalized Laplacian.
The un-normalized Laplacian ${\bf \hat{L}}$ has elements 
\bea
\hat{L}_{rq}=k_r\delta_{r,q}-{a_{rq}},
\eea
where $a_{rq}$ is the generic element of their adjacency matrix and $k_r$ is the degree of node $r\in\{1,2,\ldots,N\}$.
The normalized Laplacian $\bf L$ of a network has instead elements
\bea
{L}_{rq}=\delta_{r,q}-\frac{a_{rq}}{\sqrt{k_rk_q}}.
\eea
Their spectrum is in general distinct for non-regular networks having nodes of different degree.  However as we will observe later,  their spectral dimension is the same in the large network limit.

Here we start from the normalized Laplacian and we predict the spectral dimension analytically.
This analytical calculation will be done using the renormalization group which acts on a more general class of graphs in which the links can be weighted, so in general we are interested to study the fixed-point properties of spectrum of weighted normalized Laplacian matrices $\bf L$ of elements
\bea
L_{rq}=\delta_{rq}-\frac{w_{rq}}{\sqrt{s_rs_q}},
\eea
where $w_{rq}$ indicates the weight of link $(r,q)$ and $s_r$ indicates the strength of node  $r$, i.e. $s_r=\sum_{q=1}^N w_{rq}$.

The {\it spectral dimension} $d_S$  determines the scaling of the density of eigenvalues $\rho(\mu)$ of the normalized Laplacian of networks with distinct geometrical properties. In particular, in presence of the spectral dimension for $\mu\ll1$ we observe the asymptotic behavior
\bea
\rho(\mu)\cong C \mu^{d_S/2-1},
\label{scaling}
\eea
where $C$ is independent of $\mu$.

In $d$-dimensional Euclidean lattices the spectral dimension coincides with the Hausdorff dimension $d_S=d_H=d$. More generally, it can be shown that $d_S$ is related to the Hausdorff dimension $d_H$ of the network by the inequalities \cite{Jonsson1,Jonsson2}
\bea
d_H\geq d_S\geq 2\frac{d_H}{d_H+1}. 
\label{eq:dHdS}
\eea
It follows that  for small-world networks, having infinite Hausdorff dimension $d_H=\infty$, it is only possible to have finite spectral dimension greater or equal than two, i.e.
\bea
d_S\geq 2.
\eea

Additionally we mention here that in presence of a finite spectral dimension, the cumulative distribution $\rho_c(\mu)$ evaluating the density of eigenvalues $\mu'\leq \mu$ follows the scaling 
\bea
\rho_c(\mu)\cong \tilde{C} \mu^{d_S/2}, 
\label{eq:rho_c}
\eea
for $\mu\ll 1$.  This relation it is useful  to evaluate the spectral dimension numerically, as we will do in order to compare our analytical results with numerical results.

\subsection{Gaussian model}

In order to predict the spectral dimension of a network it is useful to consider \cite{Kahng} the corresponding Gaussian model whose  partition function  reads
\bea
Z(\mu)=\int {\mathcal D}\psi \exp\left[i\mu \sum_{r}\psi_r^2-i\sum_{rq}L_{rq}\psi_r\psi_q\right]=\frac{(i\pi)^{N_n/2}}{\sqrt{\prod_{r}(\mu-\mu_r)}}
,
\label{z}
\eea
where $\mu_r$ are the eigenvalue of the normalized Laplacian matrix ${\bf L}$ and 
\bea
{\mathcal D}\psi=\prod_{r=1}^{N_n}\left(\frac{d\psi_r}{\sqrt{2\pi}}\right).
\eea  By changing variables and putting $\phi=\psi/\sqrt{s_r}$ the  partition function can  be  rewritten as 
\bea
Z(\mu)=\prod_{r}\sqrt{s_r}\int {\mathcal D}\phi \exp\left[\sum_{(r,q)\in E}-i(1-\mu)w_{rq} (\phi_r^2+\phi^2_q)-iw_{rq}\phi_r\phi_q\right],
\eea
where $E$ indicates the set of links of the network.
The spectral density $\rho(\mu)$ of the normalized Laplacian matrix can be found using the relation 
\bea
\rho(\mu)=-\frac{2}{\pi}\mbox{Im} \frac{\partial f}{\partial \mu},
\eea 
where $f$ is the free-energy density  defined as 
\bea
f=-\lim_{n\to\infty}\frac{1}{N_n}\ln Z(\mu).
\label{rm}
\eea
In fact we can use Eq. (\ref{z}) to show that 
\bea
f=-\lim_{n\to \infty}\frac{1}{N_n}\sum_{r=1}^{N_n}\frac{1}{2}\ln(\mu-\mu_r)+\frac{1}{2}\ln (i\pi).
\eea
Using Eq.(\ref{rm}) we get 
\bea
\rho(\mu)=-\frac{2}{\pi}\mbox{Im} \frac{\partial f}{\partial \mu}=\frac{1}{\pi}\lim_{n\to \infty}\frac{1}{N_n}\mbox{Im} \sum_{r=1}^{N_n}\frac{1}{\mu-\mu_r}=\lim_{n\to \infty}\frac{1}{N_n} \sum_{r=1}^{N_n}\delta({\mu-\mu_r}).
\eea

\section{Renormalization group approach}

Under the renormalization flow, $p$ and $\mu$ are renormalized. A closer look to the problem reveals that the parameters $p$ and $\mu$ are renormalized differently for links  of different type $\ell$. Therefore we parametrize the  partition function $Z_n(\bm{\omega})$ describing the partition function of the Gaussian model over a network grown  up to iteration $n$ with the parameters $\bm{\omega}=(\{\mu_{\ell}\},\{p_{\ell}\})$, i.e.
\bea
 Z_n(\bm{\omega})=\int \mathcal{D}\phi\prod_{\ell=1}^{M}\prod_{(r,q)\in E_n^{(\ell)}}z_n^{(\ell)}(\phi_r,\phi_q),
 \eea
where $M$ indicates the total number of iterations and where
\bea
z_n^{(\ell)}(\phi_r,\phi_q)&=&\exp[-i(1-\mu_{\ell})p_{\ell}(\phi_r^2+\phi_q^2)+2ip_{\ell}\phi_r\phi_q],\nonumber \\
\eea
with $E_n^{(\ell)}$ indicating the set of links of type $\ell$ in a network evolved up to iteration $n$.
The Gibbs measure of this Gaussian model reads
\bea
P_n(\{\bm{\phi}\})=\frac{1}{Z(\bm{\omega})}\prod_{\ell=1}^{n}\prod_{(r,q)\in E_n^{(\ell)}}z_n^{(\ell)}(\phi_r,\phi_q)=\frac{1}{Z(\bm{\omega})}e^{-iH(\{\phi\})},
\label{Gibbs}
\eea
where the {\em Hamiltonian} $H(\{\phi\})$ is given by 
\bea
H(\{\phi\})=\sum_{\ell=1}^{n}\sum_{(r,q)\in E_n^{(\ell)}}\left[-(1-\mu_{\ell})p_{\ell}(\phi_r^2+\phi_q^2)+2p_{\ell}\phi_r\phi_q\right]
\eea
Let us indicate with ${\mathcal N_n}$ the nodes added at iteration $n$.
We consider the following real space renormalization group procedure to calculate the free energy of the Gaussian model. We start with initial conditions $\mu_\ell=\mu$ and $p_{\ell}=1$ for all values of $\ell>0$. At each RG iteration, we integrate over the Gaussian variables $\phi_r$ associated to nodes $r\in {\mathcal N_n}$ and we rescale the remaining Gaussian variables in order to obtain the  renormalized Gibbs measure $P(\{\bm{\phi'}\})$ of the same type as Eq.~(\ref{Gibbs}) but with rescaled parameters $(\{\mu_{\ell}^{\prime}\},\{p_{\ell}^{\prime}\})$, i.e. 
\bea
P_{n-1}(\{\bm{\phi'}\})=\left.\int {\mathcal D}\phi^{(n)} P_n(\{\bm{\phi}\})\right|_{\bm{\phi'}={\bf F}(\{\bm{\phi}\})},
\eea
where
\bea
{\mathcal D}\phi^{(n)}=\prod_{r\in {\mathcal N_n}}\left(\frac{d\phi_r}{\sqrt{2\pi}}\right).
\eea 
The rescaling of the field is done in such a way that  for a  $d$-dimensional deterministic simplicial complex, $p_{1}=1$ at each iteration of the RG flow.
Then at each step of the RG transformation we have 
\bea
H(\{\phi\})\to H^{\prime}(\{\phi^{\prime}\})
,
\eea
where 
\bea
H^{\prime}(\{\phi\})=\sum_{\ell=1}^{n-1}\sum_{(r,q)\in E_{n-1}^{(\ell)}}\left\{-(1-\mu_{\ell}^{\prime})p_{\ell}^{\prime}\left[(\phi_r^{\prime})^2+(\phi_q^{\prime})^2\right]+2p_{\ell}^{\prime}\phi_r^{\prime}\phi_q^{\prime}\right\}
.
\eea
In this way we define a  group transformation acting on the model parameters ${\bm{\omega}}=(\{\mu_{\ell}\},\{p_{\ell}\})$ so that 
\bea
\bm{\omega}'=R \bm{\omega}.
\eea
Under this renormalization flow, the partition function  transforms in the following way:  
\bea
Z_{n}(\bm{\omega})=e^{-N_n g(\bm{\omega})}Z_{n-1}(\bm{\omega}').
\label{ZRG}
\eea
Using  Eq. (\ref{NA}) and Eq. (\ref{NP}) for the number of nodes $N_n$ at iteration $n$ respectively for the Apollonian and the pseudo-fractal network, the free energy density
\bea
f&=&-\lim_{n\to \infty}\frac{1}{N_n}\ln Z_n(\bm{\omega})
\label{fg}
\eea
can be written as 
\bea
f&\simeq&\sum_{\tau=0}^{\infty}\frac{g(R^{(\tau)}\bm{\omega})}{d^{\tau}}
\label{fg1}
\eea
for the Apollonian network and as 
\bea
f&\simeq&\sum_{\tau=0}^{\infty}\frac{g(R^{(\tau)}\bm{\omega})}{(d+1)^{\tau}}
\label{fg2}
\eea
for the pseudo-fractal network.

Interestingly, we anticipate here that the RG flow will be determined by a fixed point having $\mu^{\star}=0$. This result reveals that indeed the spectral dimension here calculated for normalized Laplacian is universal, i.e. in the large network limit, the spectral dimension of the normalized Laplacian is the same as the spectral dimension of the un-normalized  Laplacian as already observed in Ref. \cite{Burioni1}.

\section{Apollonian network}

\subsection{General RG equations}

The renormalization group equations for the Apollonian networks of arbitrary topological dimension $d$ can be obtained  using the general renormalization group approach described in the previous paragraph. Therefore at each renormalization group step we need first to integrate over the fields $\phi_r$ with $r\in {\mathcal N}_n$ and subsequently perform a rescaling of the fields to guarantee that $p_1=1$ at the next iteration.
 Since any node $r\in {\mathcal N}_n$ added at generation $n$ is only connected to nodes at the previous generations the integration over the corresponding field $\phi_r$ can be done independently for any node $r\in {\mathcal N}_n$.
 
The integration over a single Gaussian variable $\phi_r$ with $r\in {\mathcal N}_n$ can be easily done and is given by  
\bea
I=\int d\phi_r\prod_{q=1}^d z^{(1)}(\phi_q,\phi_r)&=&(-\pi i/d)^{1/2}G(\mu_1)^{-1/2}\exp\left[-i(1-\mu_1)\sum_{q=1}^d\phi_q^2\right]\nonumber \\
&&\times\exp\left[i\frac{\left(\sum_{q=1}^d\phi_q\right)^2}{d(1-\mu_1)}\right],
\label{I}
\eea
where 
\bea
G(\mu)&=&1-\mu.
\eea
Of course, at each step of the RG procedure we will need to integrate over each node $r\in {\mathcal N}_n$.
Each of these integrations will contribute to the Hamiltonian $H^{\prime}(\{\phi^{\prime}\})$ by a term 
\bea
\left[\left(2\frac{1}{d(1-\mu_1)}\right)\phi_q\phi_{q^{\prime}}\right]
\eea
for any link $(q,q^{\prime})$ incident  to the $d$-simplex added at iteration $n$ and including node $r$.
Since in the Apollonian network, there are  $w_{\ell}$ simplices of iteration $n$ incident to a link added at iteration $m=n-\ell$, the overall contribution to the link is 
\bea
\left[\left(2\frac{1}{d(1-\mu_1)}\right)w_{\ell}\phi_q\phi_{q^{\prime}}\right].
\eea
If we focus on the overall contribution to the Hamiltonian proportional to  the product of the two field $\phi_{q}$ and $\phi_{q^{\prime}}$ before rescaling we get 
\bea
\left\{2\left[p_{\ell+1}+\left(\frac{1}{d(1-\mu_1)}\right)w_{\ell}\right]\phi_q\phi_{q^{\prime}}\right\}.
\eea
After rescaling of the fields $\phi_q\to \phi^{\prime}_q$  and defining the Gibbs measure over the renormalized network formed by $n-1$ generation, i.e. putting
$\ell \to \ell-1$ we should have
\bea
\left\{2\left[p_{\ell+1}+\left(\frac{1}{d(1-\mu_1)}\right)w_{\ell}\right]\phi_q\phi_{q^{\prime}}\right\}=\left\{2p_{\ell}^{\prime}\phi^{\prime}_q\phi^{\prime}_{q^{\prime}}\right\}.
\eea
Therefore in order to ensure that the value of the parameter $p_1$ remains fixed at $p_1^{\prime}=p_1=1$ after each RG iteration we need to rescale the fields by considering the  rescaled variables
\bea
\phi'&=&\phi\left[p_2+\frac{d-1}{d(1-\mu_1)}\right]^{1/2},
\label{rescaling}
\eea
where we have used $w_{1}=(d-1)$.
Finally by using Eq. (\ref{wl}) for $w_{\ell}$, the RG equation for $p^{\prime}_{\ell}$ reads 
\bea
p'_{\ell}&=&\left[p_{\ell+1}+\frac{(d-1)(d-2)^{\ell-1}}{d(1-\mu_1)}\right]\left[p_2+\frac{d-1}{d(1-\mu_1)}\right]^{-1}.
\eea
Now we will proceed similarly to find the RG equations for $\mu_{\ell}^{\prime}$.
Each integration over the Gaussian variable $\phi_r$  of a node $r\in {\mathcal N}_n$  contributes a factor 
\bea
\left[\left(-(1-\mu_1)+\frac{1}{d(1-\mu_1)}\right)\phi_q^2\right]
\eea
to the Hamiltonian
for any node $q$ belonging to the $d$-simplex added at iteration $m=n-\ell$ and including node $r$.
Since there are ${v}_{\ell}$ simplicies of iteration $n$ incident to a node added at iteration $m=n-\ell$ the integration over the Gaussian variable at iteration $n$ provides, for each node $q$, a contribution to the Hamiltonian  given by  
\bea
\left[\left(-(1-\mu_1)+\frac{1}{d(1-\mu_1)}\right){v}_{\ell}\phi_q^2\right].
\eea
By identifying the overall term of  the Hamiltonian that is proportional to $\phi_q^2$ before and after the rescaling of the fields we obtain the equation 
\bea
&&\left\{-\sum_{\ell'=1}^{\ell}(1-\mu_{\ell^{\prime}+1})p_{\ell^{\prime}+1}v_{\ell-\ell^{\prime}}+\left(-(1-\mu_1)+\frac{1}{d(1-\mu_1)}\right){v}_{\ell}(\phi^{\prime}_q)^2\right\}=\nonumber \\
&&\left\{-\sum_{\ell'=1}^{\ell}(1-\mu^{\prime}_{\ell^{\prime}})p_{\ell^{\prime}}^{\prime}v_{\ell-\ell^{\prime}}(\phi^{\prime}_q)^2\right\}.
\label{xprime}
\eea
We now make a useful  combinatorial observation and we note that the coefficient $v_{\ell}$ can be written as
\bea
v_{\ell}=\sum_{\ell^{\prime}=1}^{\ell}v_{\ell-\ell^{\prime}}c_{\ell^{\prime}}
\label{combinatorial}
\eea
where $c_{\ell}$ is given by 
\bea
c_{\ell}=(d-2)^{\ell-1}.
\eea
In fact, by substituting the explicit expression for $v_{\ell}$ Eq.  (\ref{combinatorial})  follows directly from the expression
\bea
d(d-1)^{r}=\sum_{k=0}^{r-1}d(d-1)^{r-1-k}(d-2)^{k}+d(d-2)^r.
\eea
Using Eq. (\ref{combinatorial}) and the expression for the rescaled field, Eq.~(\ref{rescaling}), in Eq. (\ref{xprime}) we get the RG equation for $\mu_{\ell}$ given by 
\bea
(1-\mu_{\ell}')p_{\ell}^{\prime}&=&\left((1-\mu_1)(d-2)^{\ell-1}+(1-\mu_{\ell+1})p_{\ell+1}-\frac{(d-2)^{\ell-1}}{d(1-\mu_1)}\right)\nonumber \\
&&\times \left[p_2+\frac{d-1}{d(1-\mu_1)}\right]^{-1}.
\eea
In summary, in this paragraph we have derived the RG equation for any $d$-dimensional Apollonian networks which we rewrite here for completeness, 
\bea
&&(1-\mu_{\ell}')p_{\ell}^{\prime}=\left[(1-\mu_1)(d-2)^{\ell-1}+(1-\mu_{\ell+1})p_{\ell+1}-\frac{(d-2)^{\ell-1}}{d(1-\mu_1)}\right]\nonumber \\&&\times\left[p_2+\frac{d-1}{d(1-\mu_1)}\right]^{-1},\nonumber \\
&&p'_{\ell}=\left[p_{\ell+1}+\frac{(d-1)(d-2)^{\ell-1}}{d(1-\mu_1)}\right]\left[p_2+\frac{d-1}{d(1-\mu_1)}\right]^{-1}.
\label{RGhd}
\eea
Under the renormalization group the partition function follows Eq. (\ref{ZRG}) with 
\bea
g(\bm{\omega})=\frac{{\mathcal N}_n}{2N_n}\ln G(\mu_1)+\frac{N_{n-1}}{2N_{n}}\ln \left[p_2+\frac{1}{d(1-\mu_1)}\right]+c,
\label{gw}
\eea
where $c$ is a constant. The first term comes directly from each integration over the variables $\phi_r$ with $r\in {\mathcal N}_n$ given by Eq. (\ref{I}) and the second term comes from the rescaling of the fields.
In the following paragraphs we will first study  the RG flows in the cases $d=2$ and $d=3$ and afterwards  we  investigate the RG flow in any arbitrary dimension $d$.

\subsection{$d=2$ Farey graph}

For $d=2$ the Apollonian network reduces to a Farey graphs and the RG Eqs. (\ref{RGhd}) simplify greatly. In fact we have 
\bea
\mu_{\ell}=\mu_2
,
\nonumber \\
p_{\ell}=p\nonumber \\
\eea
for all $\ell\geq 2$.
The renormalization group transformations read then 
\bea
&(1-\mu_1')=\left((1-\mu_1)+(1-\mu_2)p-\frac{1}{2(1-\mu_1)}\right)\left[p+\frac{1}{2(1-\mu_1)}\right]^{-1},\nonumber\\
&\mu_2'=\mu_2,\nonumber \\
&p'=p\left[p+\frac{1}{2(1-\mu_1)}\right]^{-1}.
\label{RGd2}
\eea
Under the renormalization group the partition function follows Eq. (\ref{ZRG}) with $g(\bm{\omega})$ given by Eq. (\ref{gw}).

\begin{figure}[htbp]
\begin{center}
\includegraphics[width=0.7\textwidth]{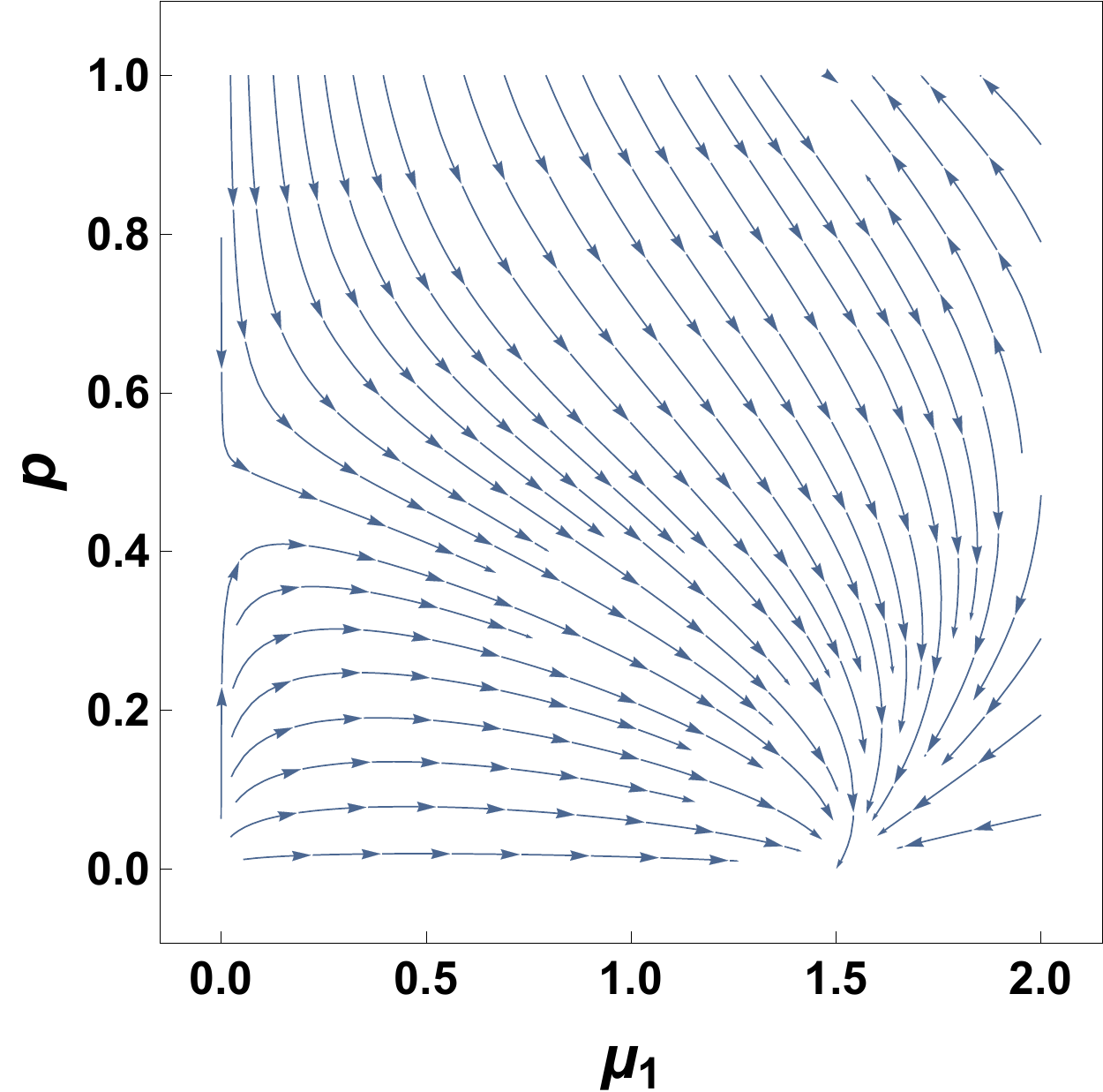}
\caption{Stream plot of the RG flow for the $d=2$ Apollonian network given by Eqs. (\ref{RGd2}) with $\mu_2=10^{-4}$.}
\label{stream_d2}
\end{center}
\end{figure}

By putting in the zero order approximation $\mu_2=\mu_2'=0$  the renormalization group equations (\ref{RGd2}) have three fixed points:
\bea
(\mu^{\star},p^{\star})&=&(0,0),\nonumber \\
(\mu^{\star},p^{\star})&=&(0,1/2),\nonumber \\
(\mu^{\star},p^{\star})&=&(3/2,0),\nonumber \\
\eea
Since we are interested in the RG flow starting from an initial condition $\mu\ll 1$ we focus on  the fixed points with $\mu^{\star}=0$.
The fixed point $(\mu,p)=(0,0)$  is unstable as it has two  eigenvalues given by $\hat{\lambda}'=4$ and $\hat{\lambda}=2$.
The fixed point $(\mu,p)=(0,1/2)$ is associated to the eigenvalues   $\lambda_1=\lambda=2$  and $\lambda_2=1/2$.
If we have initial condition $\mu_1=\mu_2=\mu$ with $0<\mu\ll1$ and $p=1$ the renormalization flow will first move toward the fixed point $(\mu,p)=(0,1/2)$ and then will move away from it along its repulsive direction.  
Close to the $(\mu^{\star},p^{\star})=(0,1/2)$ fixed point,  putting $\mu_2=\mu\ll1 $,  the linearized RG equations read
\bea
\left(\begin{array}{c}\mu_1^{\prime}\\p^{\prime}-1/2\end{array}\right)=
\left(\begin{array}{cc}2&0\\-1/4&1/2\end{array}\right)\left(\begin{array}{c}\mu_1\\p-1/2\end{array}\right)+\frac{\mu}{2}\left(\begin{array}{c}1\\0\end{array}\right)
.
\eea
At  iteration $\tau$ of the RG flow we  have
\bea
\mu^{(\tau)}&=&(\lambda^{\tau}-1)\frac{\mu}{2}
,
\\
p^{(\tau)}&=&\frac{1}{2}-\frac{1}{4}\sum_{\tau'=1}^{\tau}2^{-(\tau-\tau')} \mu^{(\tau')}+2^{-(\tau+1)}
.
\eea
Therefore for  $\tau$ large the RG flow  runs away from the   RG fixed point $(0,1/2)$ and we can approximate
\bea
\mu^{(\tau)}&\simeq&\lambda^{\tau}\frac{\mu}{2}
,
\\
p^{(\tau)}&\simeq& \frac{1}{2}-\frac{1}{6}\lambda^{\tau}\mu
.
\eea
The RG flow  is   shown in Fig.~\ref{stream_d2} where we have set initially $\mu_1=\mu_2=\mu=10^{-4}$.
Using Eq. (\ref{fg}) free-energy density can be therefore written as 
\bea
f&=&\sum_{\tau=0}^{\infty}\frac{g(R^{(\tau)}\bm{\omega})}{d^{\tau}}\nonumber \\ &&\simeq \sum_{\tau=0}^{\infty}\frac{1}{d^{\tau}}\left\{
\frac{(d-1)}{2d}\ln (1-\mu_1^{(\tau)})+\frac{1}{2d}\ln \left[p^{(\tau)}+\frac{1}{d(1-\mu_1^{(\tau)})}\right]\right\}.
\eea
Therefore we have  the spectral density $\rho(\mu)$ given by 
\bea
&&\rho(\mu)\simeq \frac{2}{\pi}\mbox{Im}\sum_{\tau=0}^{\infty}\frac{1}{d^{\tau}}\frac{\partial g(\mu_1^{\tau},1)}{\partial \mu}\nonumber \\
&&\hspace{-3mm}\simeq \frac{2}{\pi}\mbox{Im}\sum_{\tau=0}^{\infty}\frac{\lambda^{\tau}}{d^{\tau}}\left\{\frac{(d-1)}{2d}\frac{1}{1-\mu_1^{(\tau)}}+\frac{1}{2d}\frac{1}{p^{(\tau)}+1/\left[d\left(1-\mu_1^{(\tau)}\right)\right]}\left(-\frac{1}{3}+\frac{1}{d\left(1-\mu_1^{(\tau)}\right)^2}\right)\right\}.\nonumber
\eea
By approximating the sum over $\tau$ with an integral and changing the variable of integration to  $z=\lambda^{\tau}$ upon using the theorem of residues to solve the integral, we can derive the asymptotic scaling of the density of eigenvalues $\rho(\mu)$. This asymptotic scaling for $\mu\ll 1$ is  given by 
\bea
\rho(\mu)&\simeq& C \mu^{d_S/2-1}
,
\eea
where the spectral dimension $d_S$ is given by 
\bea
d_S=2\frac{\ln d}{\ln \lambda}=2.
\eea

\subsection{$d=3$ Apollonian graph}

For $d=3$  the RG Eqs. (\ref{RGhd})  simplify significantly. In fact we have 
\bea
\mu_{\ell}=\mu_1
\eea
for all $\ell\geq 1$ and 
\bea
p_{\ell}=p
\eea
for all $\ell\geq 2$.
The RG equations differ from the ones derived in the case $d=2$, and they read
\bea
&&(1-\mu_1')=\left((1-\mu_1)+(1-\mu_2)p-\frac{1}{d(1-\mu_1)}\right)\left[p+\frac{d-1}{2(1-\mu_1)}\right]^{-1},\nonumber\\
&&p'=1.
\label{RGd3}
\eea

Under the renormalization group the partition function follows Eq. (\ref{ZRG}) with $g(\bm{\omega})$ given by Eq. (\ref{gw}).
The renormalization group equations (\ref{RGd3}) give $p=1$ and reduce to a single non trivial RG equation for $\hat{\mu}=\mu_1$,
\bea
(1-\hat{\mu}^{\prime})&=&\left((1-\hat{\mu})+(1-\hat{\mu})-\frac{1}{d(1-\hat{\mu})}\right)\left[1+\frac{d-1}{d(1-\hat{\mu})}\right]^{-1},\nonumber\\
\eea
which has two fixed points:
\bea
\mu^{\star}&=&0\nonumber\\
\mu^{\star}&=&4/3.
\eea
For $\mu\ll1$ the relevant fixed point is $\mu^{\star}=0$ which has a non-trivial associated eigenvalue given by $\lambda=9/5$.
Therefore under the  RG flow we have  that at iteration $\tau$ of the RG flow
\bea
(\mu_1^{(\tau)},p^{(\tau)})=(\lambda^{\tau}\mu,1)
\eea 
with $\mu$ indicating the initial condition $\mu=\mu_1^{(1)}$.
Using Eq. (\ref{fg}) free-energy density can be therefore written as 
\bea
f&=&\sum_{\tau=0}^{\infty}\frac{g(R^{(\tau)}\bm{\omega})}{d^{\tau}}\nonumber \\ &&\simeq \sum_{\tau=0}^{\infty}\frac{1}{d^{\tau}}\left\{
\frac{(d-1)}{2d}\ln (1-\mu_1^{(\tau)})+\frac{1}{2d}\ln \left[1+\frac{d-1}{d(1-\mu_1^{(\tau)})}\right]\right\}.
\eea
Therefore we have the spectral density $\rho(\mu)$  given by 
\bea
&&\rho(\mu)\simeq \frac{2}{\pi}\mbox{Im}\sum_{\tau=0}^{\infty}\frac{1}{d^{\tau}}\frac{\partial g(\mu_1^{\tau},1)}{\partial \mu}\nonumber \\
&&\simeq\frac{2}{\pi}\mbox{Im}\sum_{\tau=0}^{\infty}\frac{\lambda^{\tau}}{d^{\tau}}\left\{\frac{(d-1)}{2d}\frac{1}{1-\mu_1^{(\tau)}}+\frac{1}{2d}\frac{1}{ \left[d\left(1-\mu_1^{(\tau)}\right)+{d-1}\right]}\frac{d-1}{\left(1-\mu_1^{(\tau}\right)}\right\}
.
\nonumber
\eea
By proceeding similarly to the case $d=2$ and approximating the sum over $\tau$ with an integral, we obtain the asymptotics 
\bea
\rho(\mu)&\cong& C \mu^{d_s/2-1}
\eea
valid for $\mu\ll 1$
with the spectral dimension $d_s$  given by 
\bea
d_s=2\frac{\ln d}{\ln \lambda}=2\frac{\ln d}{\ln 9/5}=3.73813\ldots.
\eea
Interestingly, Apollonian networks in $d=3$ are planar. As we will see when comparing the spectral dimension of Apollonian network with the spectral dimension of NGFs, the randomness introduced by the NGF constructions always lower the spectral dimension of the network.

\subsection{$d>3$ dimensional Apollonian graph}

Let us now determine the RG flow in the general case of a $d$-dimensional Apollonian network.
By putting 
\bea
x_{\ell}=(1-\mu_{\ell})p_{\ell}
,
\label{xell}
\eea
the RG Eqs. (\ref{RGhd})  relating the parameters $(\{x_{\ell}^{(\tau)}\},\{p_{\ell}^{(\tau)}\})$ at iteration $\tau$ of the RG flow with the parameters $(\{x_{\ell}^{(\tau+1)}\},\{p_{\ell}^{(\tau+1)}\})$ at iteration $\tau+1$ of the RG flow read
\bea
&&x_{\ell}^{(\tau+1)}=\left[x_{\ell}^{(\tau)}+\left(x_1^{(\tau)}-\frac{1}{dx_1^{(\tau)}}\right)(d-2)^{\ell-1}\right]\left[p_2^{(\tau)}+\frac{d-1}{dx_1^{(\tau)}}\right]^{-1},\nonumber \\
&&p_{\ell}^{(\tau+1)}=\left[p_{\ell+1}^{(\tau)}+\frac{(d-1)(d-2)^{\ell-1}}{dx_1^{(\tau)}}\right]\left[p_2^{(\tau)}+\frac{d-1}{dx_1^{(\tau)}}\right]^{-1}.
\label{RGhd3}
\eea

In order to solve these equations we use the auxiliary variable $y_1^{(\tau)}$ defined as   
\bea
{y}_{1}^{(\tau+1)}=p_{2}^{(\tau+1)}+\frac{d-1}{dx_1^{(\tau+1)}}.
\eea 
The explicit solution of the RG equations (\ref{RGhd3})   equations read
\bea
p_{2}^{(\tau+1)}&=&\prod_{m=1}^{\tau}\frac{1}{y_1^{(m)}}+\frac{d-1}{d}\sum_{m=1}^{\tau}\frac{(d-2)^{\tau-m+1}}{x^{(m)}}\prod_{m'=m}^{\tau}\frac{1}{y_1^{(m')}},\nonumber\\ 
y_1^{(\tau+1)}&=&p_2^{(\tau+1)}+\frac{(d-1)}{d x_1^{(\tau+1)}}\nonumber\\
&=&\prod_{m=1}^{\tau}\frac{1}{y_1^{(m)}}+\frac{d-1}{d}\sum_{m=1}^{\tau}\frac{(d-2)^{\tau-m+1}}{x^{(m)}}\prod_{m'=m}^{\tau}\frac{1}{y_1^{(m')}}+\frac{(d-1)}{d x_1^{(\tau+1)}},\nonumber \\
x_{1}^{(\tau+1)}&=&x_1^{(1)}\prod_{m=1}^{\tau}\frac{1}{y_1^{(m)}}+\sum_{m=1}^{\tau}\left(x_1^{(m)}-\frac{1}{dx_1^{(m)}}\right)(d-2)^{\tau-m}\prod_{m'=m}^{\tau}\frac{1}{y_1^{(m')}}.\nonumber 
\eea
This explicit solution reveals that $p_2^{(\tau+1)},y_1^{(\tau+1)}$ and $x_1^{(\tau+1)}$ depend on the dimension $d$ and on all the values that the parameters $p_2^{(\tau')},y_1^{(\tau')}$ and $x_1^{(\tau')}$ take for $\tau'\leq \tau$.
However if we introduce a set of of additional auxiliary variables called $A^{(\tau)},B^{(\tau)}$ and $C^{(\tau)}$ we can express   the variables $p_2^{(\tau+1)},y_1^{(\tau+1)}$ and $x_1^{(\tau+1)}$ only as functions of the value of the additional auxiliary variable at time $\tau$. In fact we have
\bea
y_1^{(\tau+1)}&=&A^{(\tau)}+B^{(\tau)},\nonumber\\
x_{1}^{(\tau+1)}&=&(1-\mu)A^{(\tau)}+C^{(\tau)},
\label{x1y1Ap}
\eea
where  $A^{(\tau)},B^{(\tau)}$ and $C^{(\tau)}$ are defined as 
\bea
A^{(\tau)}&=&\prod_{m=1}^{\tau}\frac{1}{y_1^{(m)}},\nonumber \\
B^{(\tau)}&=&\frac{d-1}{d}\sum_{m=1}^{\tau}\frac{(d-2)^{\tau-m+1}}{x^{(m)}}\prod_{m'=m}^{\tau}\frac{1}{y_1^{(m')}}+\frac{(d-1)}{d x_1^{(\tau+1)}},
\nonumber \\C^{(\tau)}&=&\sum_{m=1}^{\tau}\left(x_1^{(m)}-\frac{1}{dx_1^{(m)}}\right)(d-2)^{\tau-m}\prod_{m'=m}^{\tau}\frac{1}{y_1^{(m')}}.\nonumber \\
\eea
The RG equations can be then solved by writing the recursive equations for $A^{(\tau)},B^{(\tau)}$ and $C^{(\tau)}$ that read 
\bea
A^{(\tau+1)}&=&\frac{1}{y_1^{(\tau+1)}}A^{(\tau)}
,
\nonumber \\
B^{(\tau+1)}&=&\frac{d-2}{y_1^{(\tau+1)}}B^{(\tau)}+\frac{d-1}{d}\frac{1}{x_1^{(\tau+2)}}
,
\nonumber \\
C^{(\tau+1)}&=&\frac{(d-2)}{y_1^{(\tau+1)}}C^{(\tau)}+\frac{1}{y_1^{(\tau+1)}}\left(x_1^{(\tau+1)}-\frac{1}{dx_1^{(\tau+1)}}\right)
.
\eea
By using Eq. (\ref{x1y1Ap}) this set of equations can be written as a closed set of equations for $A^{(\tau)},B^{(\tau)}$ and $C^{(\tau)}$ only, i.e.
\bea
A^{(\tau+1)}&=&\frac{1}{A^{(\tau)}+B^{(\tau)}}A^{(\tau)}
,
\nonumber \\
B^{(\tau+1)}&=&\frac{d-2}{A^{(\tau)}+B^{(\tau)}}B^{(\tau)}+\frac{d-1}{d}(A^{(\tau)}+B^{(\tau)})
\nonumber \\
&&\times \left[2(1-\mu)A^{(\tau)}+(d-1)C^{(\tau)}-\frac{1}{d}\frac{1}{(1-\mu)A^{(\tau)}+C^{(\tau)}}\right]^{-1}
,
\nonumber \\
C^{(\tau+1)}&=&\frac{1}{A^{(\tau)}+B^{(\tau)}}\left((d-1)C^{(\tau)}+(1-\mu)A^{(\tau)}-\frac{1}{d[(1-\mu)A^{(\tau)}+C^{(\tau)}]}\right)\nonumber 
\eea
with initial conditions $A^{(0)}=1,B^{(0)}=(d-1)/(d[1-\mu)]),C^{(0)}=0$.

The relevant fixed point of these RG equations  for $\mu=0$ is 
\bea
A^{\star}=0
,
\nonumber \\
B^{\star}=\frac{d^2-d-1}{d}
,
\nonumber \\
C^{\star}=1
.
\eea
If we consider the Jacobian matrix of the RG tranformation we get the three eigenvalues $\lambda_1>\lambda_2>\lambda_3$ given by 
\bea
\lambda_1=\frac{d^2}{d^2-d-1}
,
\nonumber \\
\lambda_2=\frac{d}{d^2-d-1}
,
\nonumber \\
\lambda_3=0
\eea
with $\lambda_1>1$ and $\lambda_2<1$.
The right eigenvectors corresponding to these eigenvalues are
\bea
{\bf u_1}=\frac{1}{c_1}\left(d+1,-d,d^2-d+1\right),\nonumber \\
{\bf u_2}=\left(1,0,0\right),\nonumber \\
{\bf u_3}=\frac{1}{c_3}\left(d^2-d-1,d,d-1\right),\nonumber \\
\eea
where $c_1$ and $c_3$ are normalization constants.
The left eigenvectors corresponding to these eigenvalues are
\bea
{\bf v_1}=\frac{1}{d_1}\left(0,d+1,-d\right),\nonumber \\
{\bf v_2}=\frac{1}{d_2}\left(-1,d-2,1\right),\nonumber \\
{\bf v_3}=\frac{1}{d_3}\left(0,d^2-d+1,d\right),\nonumber \\
\eea
where $d_1,d_2,d_3$ are normalization constants.
We call  ${\bf X}^{(\tau)}$ the column vector 
\bea
{\bf X}^{(\tau)}=\left(A^{(\tau)},B^{(\tau)},C^{(\tau)}\right)
.
\eea
Then near the fixed point ${\bf X}^{\star}$ given by 
\bea
{\bf X}^{\star}=\left(A^{\star},B^{\star},C^{\star}\right)
,
\eea
we have 
\bea
{\bf X}^{(\tau)}={\bf X}^{\star}+\sum_{m=1}^3\lambda_m^{\tau}{\bf v}_m\avg{{\bf u}_m,{\bf X}^{(0)}-{\bf X}^{\star}}.
\eea
To the leading term, we have 
\bea
{\bf X}^{(\tau)}={\bf X}^{\star}+\lambda_1^{\tau}{\bf v}_1\avg{{\bf u}_1,{\bf X}^{(0)}-{\bf X}^{\star}}
,
\eea
where the scalar product
\bea
\avg{{\bf u}_1,{\bf X}^{(0)}-{\bf X}^{\star}}\propto (d-1)\frac{\mu}{(1-\mu)}
.
\eea
Under the renormalization group transformations the partition function follows Eq. (\ref{ZRG}) with $g(\bm{\omega})$ given by Eq. (\ref{gw}).
We calculate  the leading terms of the density of eigenvalues $\rho(\mu)$ following  similar steps used in the case $d=2$ and $d=3$, and find 
\bea
\rho(\mu)\simeq C\mu^{d_S/2-1}
\eea
for $\mu\ll1$ with the spectral dimension $d_S$ given by 
\bea
d_s=2\frac{\ln d}{\ln \lambda}=-2\frac{\ln d}{\ln [1-1/d-1/d^2)]}.
\eea
For large $d$, the spectral dimension $d_S$ scales as
\bea
d_s\simeq 2\ln (d)\left[d-\frac{3}{2}+O(1/d)\right].
\eea
Therefore the spectral dimension grows with the topological dimension faster than linearly.

\subsection{Comparison to NGF with $s=-1$}

\begin{figure}[htbp]
\begin{center}
\includegraphics[width=0.99\textwidth]{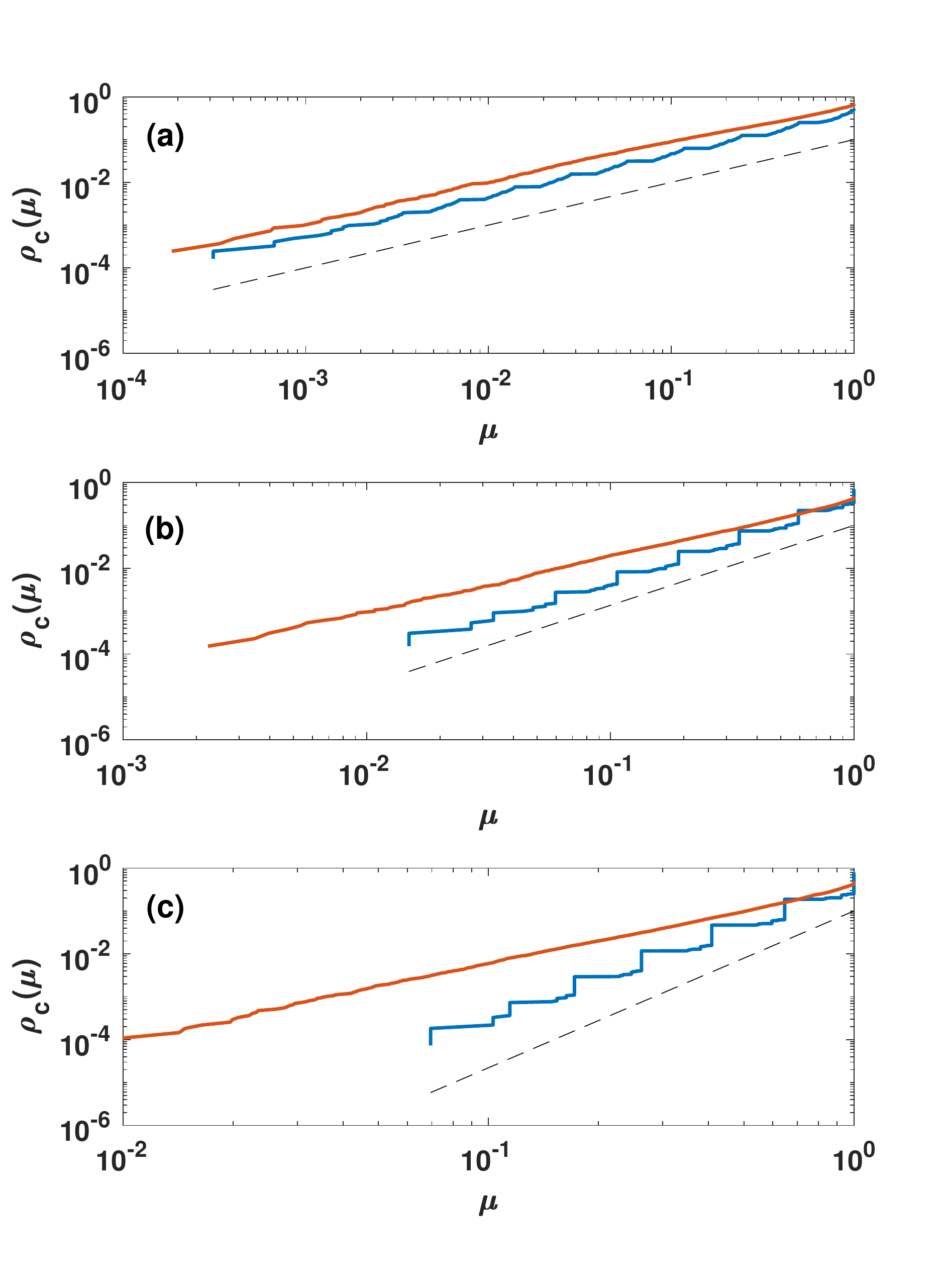}
\caption{Cumulative distribution of the eigenvalues $\rho_c(\mu)$ for the Apollonian network in $d=2$(panel a), $d=3$ (panel b) and $d=4$ (panel c) (shown in blue) are compared with the theoretical predictions of the spectral dimension (black dashed line) and with the cumulative distribution of the eigenvalues for the NGF with flavor $s=-1$ (red lines).}
\label{Apollonian_NGF}
\end{center}
\end{figure}

In this paragraph we conduct the numerical check of our analytical predictions for the spectral dimension of Apollonian networks and we compare the spectral dimension of the Apollonian network with the numerically observed spectral dimension in NGF with flavor $s=-1$ which are also called random Apollonian networks.
In Fig.~\ref{Apollonian_NGF} we compare the spectrum of the Apollonian network of dimension $d=2,d=3$ and $d=4$ to the analytical predictions for the spectral dimension finding very good agreement. Additionally we compare the spectrum of the Apollonian network of dimension $d=2,d=3$ and $d=4$ to   the spectrum of NGFs with flavor $s=-1$ and dimensions $d=2,3$ and $d=4$.
From our numerical study of the spectrum of the Apollonian networks and the NGF with $s=-1$ we draw a series of observations.
\begin{itemize}
\item
Apollonian networks are highly symmetrical structures which implies that the spectrum has many degeneracies that  complies with a finite spectral dimension.The randomness present in NGFs reduces the relevance of these degeneracies of eigenvalues as the symmetries are not exact any more for these random structures.
\item
Any randomness in the generation of the manifolds present in the NGFs can significantly change the spectral dimension. In particular it seems always to lower the spectral dimension. This is in agreement with the intuition that a random walker moving on a hyperbolic manifold like the Apollonian network will "experience" a higher spectral dimension than a random walker moving on a subgraph of this manifold, such as the NGFs with flavor $s=-1$.
\end{itemize}
It would be interesting to have also a RG approach to predict the spectral dimension of NGF. However the extension of the RG approach to random and disordered systems is  challenging and it has been so far addressed in the literature only in rare cases (see for instance Ref. \cite{Samukhin_RG}).

\section{Pseudo-fractal networks}

\subsection{RG equations}

In a pseudo-fractal networks evolved until generation $n$, we indicate as type $\ell$ all links added at iteration $m=n-\ell$.
At each iteration $n$ we add a new $d$-dimensional simplex to every $(d-1)$-dimensional face present at the previous iteration, i.e. we attach a new $d$-dimensional simplex to every face added at any iteration $m<n$. Therefore the 
 RG equation can be written directly starting from the equations valid for the Apollonian graphs by  taking into account  that each link of type $\ell$ receives the sum of the contributions coming from the integration of the Gaussian variables associated to nodes added at the last generation.
This considerations lead immediately to the RG equations
\bea
&&x_{\ell}^{\prime}=\left[x_{\ell+1}+\left(x_1-\frac{1}{d x_1}\right){\sum_{\ell'=0}^{\ell-1}(d-2)^{\ell'}}\right]\left[p_2+\frac{d-1}{dx_1}\right]^{-1},\nonumber \\
&&p'_{\ell}=\left[p_{\ell+1}+\frac{(d-1)\sum_{\ell'=0}^{\ell-1}(d-2)^{\ell'}}{dx_1}\right]\left[p_2+\frac{d-1}{dx_1}\right]^{-1},
\label{PFhd1}
\eea
where $x_{\ell}$ are defined in Eq. (\ref{xell}), $x_\ell =(1-\mu_{\ell})p_{\ell}$. 
The free energy is given by Eq.  (\ref{fg2}) where by using  a procedure similar to the one used to derive  the corresponding expression for the Apollonian network we  easily find 
\bea
g(\bm{\omega})=\frac{{\mathcal N}_n}{2N_n}\ln G(\mu_1)+\frac{N_{n-1}}{2N_{n}}\ln \left[p_2+\frac{1}{d(1-\mu_1)}\right]+c,
\label{gw2}
\eea
where $c$ is a constant.
In the following paragraphs we will study the RG flow and predict the spectral dimension for pseudo-fractal networks for dimension $d=2,d=3$ and $d>3$.

\subsection{Case $d=2$}

For $d=2$ the RG Eqs.(\ref{PFhd1}) for the pseudo-fractal network simplify significantly.  We have 
\bea
\mu_{\ell}=\mu_1
\eea
for $\ell\geq 1$ and 
\bea
p_{\ell}=p\nonumber \\
\eea
for all $\ell\geq 2$.
In particular,  
the RG equations reduce to the same equations found for the Apollonian network in dimension $d=3$  but with the difference  in the value of $d$. 
The resulting equations are
\bea
&&(1-\mu_1')=\left((1-\mu_1)+(1-\mu_2)p-\frac{1}{d(1-\mu_1)}\right)\left[p+\frac{d-1}{2(1-\mu_1)}\right]^{-1},\nonumber\\
&&p'=1.
\label{PSRGd2}
\eea
The fixed point is $\mu^{\star}_1=0$ and $p^{\star}=1$ with eigenvalue $\lambda=2$. 
A straightforward calculation of the leading term of the density of eigenvalues for $\mu\ll1$ leads to 
the 
 spectral dimension 
\bea
d_s=2\frac{\ln (d+1)}{\ln \lambda}=2\frac{\ln 3}{\ln 2}=3.16993\ldots.
\eea
We note here that the pseudo-fractal networks of dimension $d=2$ are planar. However here they are found to have a smaller spectral dimension of the planar Apollonian networks in dimension $d=3$.
One might naively think that the spectral dimension of the $d=3$ Apollonian network is the largest spectral dimension of planar networks. However a slight modification of the $d=2$ pseudo-fractal network in which at each iteration every link is attached to $k$ new triangles gives a spectral dimension that that diverges for $k\to \infty$. So planar networks can have unbounded spectral dimension.

\subsection{Case $d=3$}

In the case $d=3$ the RG Eqs.(\ref{PFhd1}) relating the parameter values at iteration $\tau+1$ with the parameter values at iteration $\tau$ are given by  
\bea
&&x_{\ell}^{(\tau+1)}=\left[x_{\ell+1}^{(\tau)}+\left(x_1^{(\tau)}-\frac{1}{d x_1^{(\tau)}}\right)\ell\right]\left[p_2^{(\tau)}+\frac{d-1}{dx_1^{(\tau)}}\right]^{-1},\nonumber \\
&&p_{\ell}^{(\tau+1)}=\left[p_{\ell+1}^{(\tau)}+\frac{(d-1)\ell}{dx_1^{(\tau)}}\right]\left[p_2^{(\tau)}+\frac{d-1}{dx_1^{(\tau)}}\right]^{-1},
\label{PFhd3}
\eea
where $x_{\ell}$ is defined in Eq. (\ref{xell}). In fact these equations can be directly derived from Eqs. (\ref{PFhd1}) by performing the sum over $\ell'$ in the case $d=3$.
By defining the auxiliary variable 
\bea
y_1^{(\tau+1)}&=&p_2^{(\tau+1)}+\frac{d-1}{d}\frac{1}{x_1^{(\tau+1)}}
,
\eea
we can express the explicit solution of the RG equations (\ref{PFhd3}) as
\bea
x_1^{(\tau+1)}&=&(1-\mu)\prod_{m=1}^{\tau}\frac{1}{y_1^{(m)}}+\sum_{m=1}^{\tau}\left(x_1^{(m)}-\frac{1}{dx_1^{(m)}}\right)(\tau+1-m)\prod_{m'=m}^{\tau}\frac{1}{y^{(m')}}
,
\nonumber \\
p^{(\tau+1)}_2&=&\prod_{m=1}^{\tau}\frac{1}{y_1^{(m)}}+\frac{d-1}{d}\sum_{m=1}^{\tau}\frac{1}{x_1^{(m)}}(\tau+2-m)\prod_{m'=m}^{\tau}\frac{1}{y^{(m')}}.
\eea
We now write this expression in terms of the auxiliary variables  $A^{(\tau)},B^{(\tau)},C^{(\tau)}$ as
\bea
y_1^{(\tau+1)}&=&A^{(\tau)}+B^{(\tau)}+\frac{d-1}{d}\frac{1}{x_1^{(\tau+1)}}
,
\nonumber \\
x_1^{(\tau+1)}&=&(1-\mu)A^{(\tau)}+C^{(\tau)}
,
\eea
where we have put 
\bea
A^{(\tau)}&=&\prod_{m=1}^{\tau}\frac{1}{y_1^{(m)}},\nonumber \\
B^{(\tau)}&=&\frac{d-1}{d}\sum_{m=1}^{\tau}\frac{1}{x_1^{(m)}}(\tau+2-m)\prod_{m'=m}^{\tau}\frac{1}{y^{(m')}},\nonumber\\
C^{(\tau)}&=&\sum_{m=1}^{\tau}\left(x_1^{(m)}-\frac{1}{dx_1^{(m)}}\right)(\tau+1-m)\prod_{m'=m}^{\tau}\frac{1}{y^{(m')}}.
\label{def1Ad3}
\eea
In this case it is impossible to write a recursive equation for $A^{(\tau+1)},B^{(\tau+1)},C^{(\tau+1)}$ depending only on the variables $A^{(\tau)},B^{(\tau)},C^{(\tau)}$. It is possible however to circumvent this difficulty by defining a further pair of auxiliary variables $D^{(\tau)}$ and $E^{(\tau)}$ given by 
\bea
D^{(\tau)}&=&+\sum_{m=1}^{\tau}\left(x_1^{(m)}-\frac{1}{dx_1^{(m)}}\right)\prod_{m'=m}^{\tau}\frac{1}{y^{(m')}},\nonumber \\
E^{(\tau)}&=&\frac{d-1}{d}\sum_{m=1}^{\tau}\frac{1}{x_1^{(m)}}\prod_{m'=m}^{\tau}\frac{1}{y^{(m')}}.
\label{def2Ad3}
\eea
In this way we can study the RG flow by studying the behavior of the following set of recursive equations close to their relevant fixed point. These equations read
\bea
x_1^{(\tau+1)}&=&(1-\mu)A^{(\tau)}+C^{(\tau)}
,
\nonumber \\
y_1^{(\tau+1)}&=&A^{(\tau)}+B^{(\tau)}+\frac{d-1}{d}\frac{1}{[(1-\mu)A^{(\tau)}+C^{(\tau)}]}
,
\nonumber \\
A^{(\tau+1)}&=&\frac{1}{y_1^{(\tau+1)}}A^{(\tau)}
,
\nonumber \\
B^{(\tau+1)}&=&\frac{1}{y_1^{(\tau+1)}}\left[B^{(\tau)}+E^{(\tau)}+2\frac{d-1}{d}\frac{1}{x_1^{(\tau+1)}}\right]
,
\nonumber \\
C^{(\tau+1)}&=&\frac{1}{y_1^{(\tau+1)}}\left[C^{(\tau)}+D^{(\tau)}+\left(x_1^{(\tau+1)}-\frac{1}{dx_1^{(\tau+1)}}\right)\right]
,
\nonumber \\
D^{(\tau+1)}&=&\frac{1}{y_1^{(\tau+1)}}\left[D^{(\tau)}+\left(x_1^{(\tau+1)}-\frac{1}{dx_1^{(\tau+1)}}\right)\right]
,
\nonumber \\
E^{(\tau+1)}&=&\frac{1}{y_1^{(\tau+1)}}\left[E^{(\tau)}+\frac{d-1}{d}\frac{1}{x_1^{(\tau+1)}}\right],
\label{RGd3psA}
\eea
with initial conditions 
$A^{(1)},B^{(1)},C^{(1)},D^{(1)},E^{(1)}$ which can be found by inserting in Eqs. (\ref{def1Ad3}) and (\ref{def2Ad3}) $x_1^{(1)}=1-\mu$ and $y_1^{(1)}=[1+\frac{d+1}{d(1-\mu)}]$.
The relevant fixed point of these equations is 
\bea
A^{\star}&=&0
,
\nonumber \\
B^{\star}&=&\frac{1}{6}\left(1+3+\sqrt{28}\right)
,
\nonumber \\
C^{\star}&=&1
,
\nonumber \\
D^{\star}&=&1/2\left(-1+\frac{1}{3}+\frac{\sqrt{28}}{3}\right)
,
\nonumber \\
E^{\star}&=&1/2\left(-1+\frac{1}{3}+\frac{\sqrt{28}}{3}\right)
.
\nonumber \\
\eea
Close to this fixed point, the RG equations (\ref{RGd3psA}) have the relevant eigenvalue
\bea
\lambda=1.68471
.
\eea
All other eigenvalues are real  non-negative and smaller than one. Therefore they are negligible.
By performing the study of the density of eigenvalues $\rho(\mu)$ for $\mu\ll 1$ we can derive the value of the spectral dimension $d_s$ given by 
\bea
d_s=2\frac{\ln(d+1)}{\ln\lambda}=5.31562\ldots.
\eea

\subsection{Case $d>3$}

\begin{figure}[htbp]
\begin{center}
\includegraphics[width=0.7\textwidth]{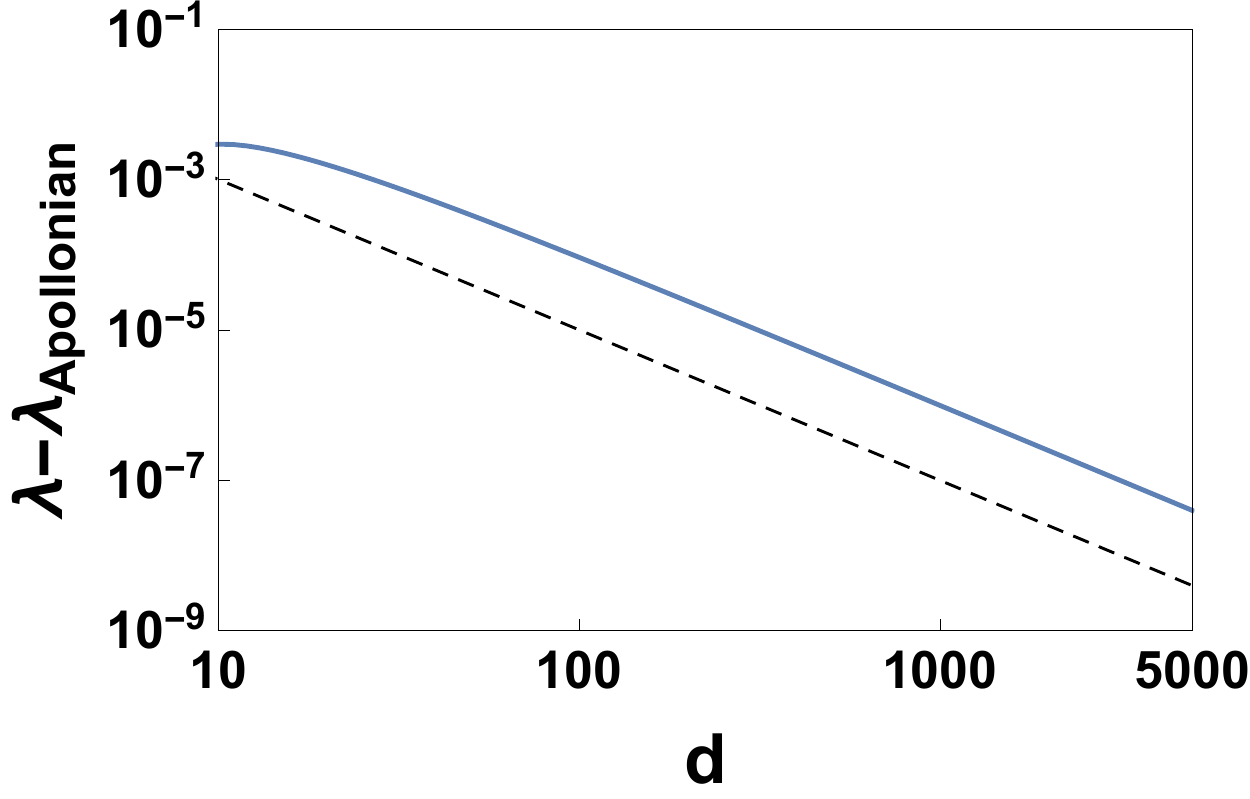}
\caption{The difference between the leading eigenvalue $\lambda$ of the RG flow of the pseudo-fractal network and the leading eigenvalue $\lambda_{\mbox{Apollonian}}=d^2/(d^2-d-1)$  is plotted versus the topological dimension $d$ (solid line). The dashed line with slope $-2$ is a guide to the eye. }
\label{lambdadiff}
\end{center}
\end{figure}

\begin{figure}[htbp]
\begin{center}
\includegraphics[width=0.7\textwidth]{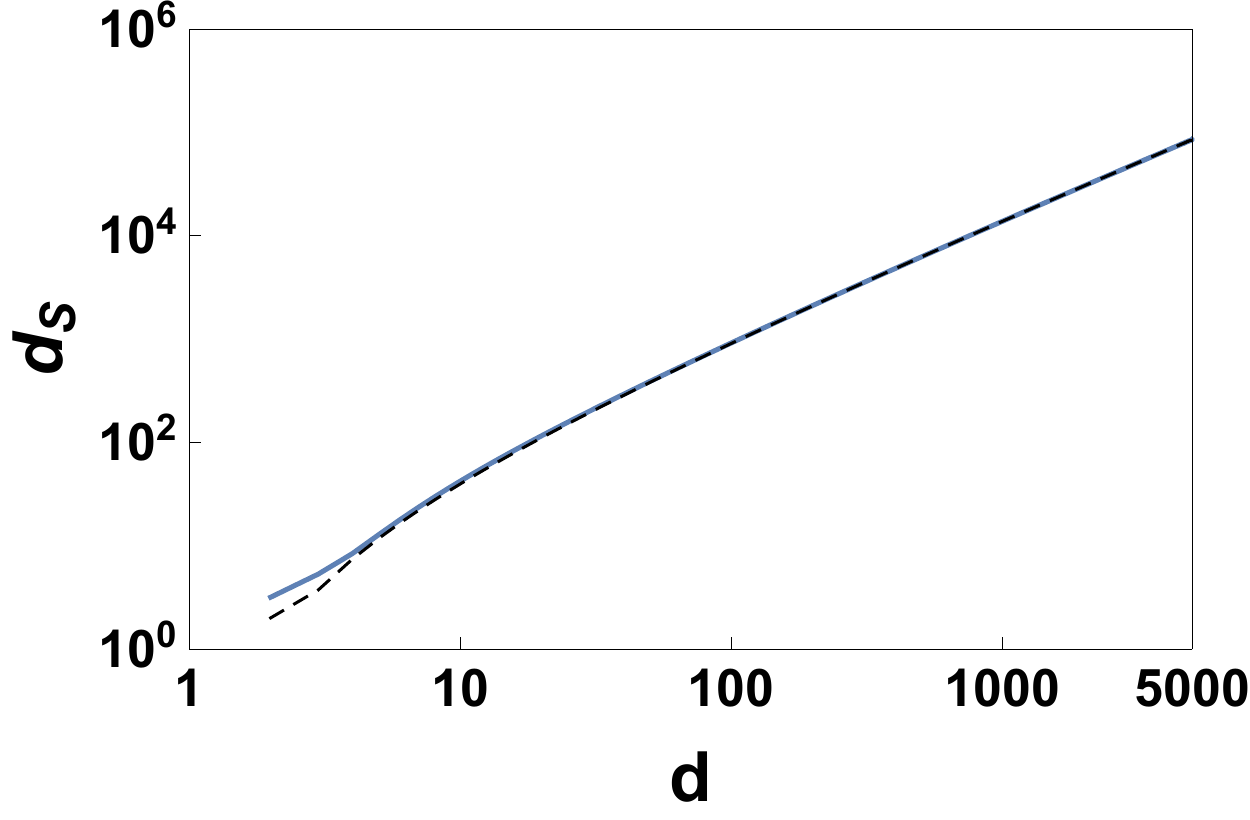}
\caption{The spectral dimension $d_s$ of the $d$-dimensional pseudo-fractal network calculated numerically from the RG equations is plotted versus $d$ (solid line) and compared with the spectral dimension of the $d$-dimensional Apolloanian networks (dashed line).}
\label{dsspeudo}
\end{center}
\end{figure}

For the  pseudo-fractal network of dimension $d>3$, the RG Eqs. (\ref{PFhd1}) relating the parameter values at the iteration $\tau+1$ of  the RG flow to the parameter values at iteration $\tau$, are given by 
\bea
&&x_{\ell}^{(\tau+1)}=\left[x_{\ell+1}^{(\tau)}+\left(x_1^{(\tau)}-\frac{1}{d x_1^{(\tau)}}\right)\frac{[(d-2)^{\ell}-1]}{d-3}\right]\left[p_2^{(\tau)}+\frac{d-1}{dx_1^{(\tau)}}\right]^{-1},\nonumber \\
&&p^{(\tau+1)}_{\ell}=\left[p_{\ell+1}^{(\tau)}+\frac{(d-1)}{dx_1^{(\tau)}}\frac{[(d-2)^{\ell}-1]}{d-3}\right]\left[p_2^{(\tau)}+\frac{d-1}{dx_1^{(\tau)}}\right]^{-1},
\label{RGhd}
\eea
where $x_{\ell}$ is defined in Eq. (\ref{xell}).
In order to solve these equations we put   
\bea
{y}_{1}^{(\tau+1)}=p_{2}^{(\tau+1)}+\frac{d-1}{dx_1^{(\tau+1)}}.
\eea 
The  solution of the Eqs. (\ref{RGhd}) reads
\bea
p_{2}^{(\tau+1)}&=&\prod_{m=1}^{\tau}\frac{1}{y_1^{(m)}}+\frac{(d-1)}{d(d-3)}\sum_{m=1}^{\tau}\frac{[(d-2)^{\tau-m+2}-1]}{x^{(m)}}\prod_{m'=m}^{\tau}\frac{1}{y_1^{(m')}}
,
\nonumber\\ 
y_1^{(\tau+1)}&=&p_2^{(\tau+1)}+\frac{(d-1)}{d x_1^{(\tau+1)}}\nonumber\\
&=&\prod_{m=1}^{\tau}\frac{1}{y_1^{(m)}}+\frac{(d-1)}{d(d-3)}\sum_{m=1}^{\tau}\frac{[(d-2)^{\tau-m+2}-1]}{x^{(m)}}\prod_{m'=m}^{\tau}\frac{1}{y_1^{(m')}}+\frac{(d-1)}{d x_1^{(\tau+1)}}
,
\nonumber \\
x_{1}^{(\tau+1)}&=&x_1^{(1)}\prod_{m=1}^{\tau}\frac{1}{y_1^{(m)}}+\frac{1}{d-3}\sum_{m=1}^{\tau}\left(x_1^{(m)}-\frac{1}{dx_1^{(m)}}\right)[(d-2)^{\tau+1-m}-1]\prod_{m'=m}^{\tau}\frac{1}{y_1^{(m')}}
.
\nonumber 
\eea
Let us now put
\bea
y_1^{(\tau+1)}&=&A^{(\tau)}+B^{(\tau)}-D^{(\tau)}+\frac{d-1}{dx_1^{(\tau+1)}}
,
\nonumber\\
x_{1}^{(\tau+1)}&=&(1-\mu)A^{(\tau)}+C^{(\tau)}-E^{(\tau)}\nonumber 
\eea
with $A^{(\tau)},B^{(\tau)},C^{(\tau)},D^{(\tau)},E^{(\tau)}$ given by 
\bea
A^{(\tau)}&=&\prod_{m=1}^{\tau}\frac{1}{y_1^{(m)}}
,
\nonumber \\
B^{(\tau)}&=&=\frac{(d-1)}{d(d-3)}\sum_{m=1}^{\tau}\frac{(d-2)^{\tau-m+2}}{x^{(m)}}\prod_{m'=m}^{\tau}\frac{1}{y_1^{(m')}}
,
\nonumber \\
C^{(\tau)}&=&\frac{1}{d-3}\sum_{m=1}^{\tau}\left(x_1^{(m)}-\frac{1}{dx_1^{(m)}}\right)(d-2)^{\tau+1-m}\prod_{m'=m}^{\tau}\frac{1}{y_1^{(m')}}
,
\nonumber \\
D^{(\tau)}&=&\frac{(d-1)}{d(d-3)}\sum_{m=1}^{\tau}\frac{1}{x^{(m)}}\prod_{m'=m}^{\tau}\frac{1}{y_1^{(m')}}
,
\nonumber \\
E^{(\tau)}&=&\frac{1}{d-3}\sum_{m=1}^{\tau}\left(x_1^{(m)}-\frac{1}{dx_1^{(m)}}\right)\prod_{m'=m}^{\tau}\frac{1}{y_1^{(m')}}.
\label{defAdg3}
\eea
The RG flow can be cast in a set of recursive equations for $A^{(\tau)},B^{(\tau)},C^{(\tau)},D^{(\tau)},E^{(\tau)}$. In fact we have
\bea
y_1^{(\tau+1)}&=&A^{(\tau)}+B^{(\tau)}-D^{(\tau)}+\frac{d-1}{d[(1-\mu)A^{(\tau)}+C^{(\tau)}-E^{(\tau)}]}
,
\nonumber\\
x_{1}^{(\tau+1)}&=&(1-\mu)A^{(\tau)}+C^{(\tau)}-E^{(\tau)}
,
\nonumber \\
A^{(\tau+1)}&=&\frac{1}{y_1^{(\tau+1)}}A^{(\tau)}
,
\nonumber \\
B^{(\tau+1)}&=&\frac{d-2}{y_1^{(\tau+1)}}B^{(\tau)}+\frac{(d-1)(d-2)^2}{d(d-3)}\frac{1}{y_1^{(\tau+1)}x_1^{(\tau+1)}}
,
\nonumber \\
C^{(\tau+1)}&=&\frac{(d-2)}{y_1^{(\tau+1)}}C^{(\tau)}+\frac{(d-2)}{(d-3)}\frac{1}{y_1^{(\tau+1)}}\left(x_1^{(\tau+1)}-\frac{1}{dx_1^{(\tau+1)}}\right)
,
\nonumber \\
D^{(\tau+1)}&=&\frac{1}{y_1^{(\tau+1)}}D^{(\tau)}+\frac{(d-1)}{d(d-3)}\frac{1}{y_1^{(\tau+1)}x_1^{(\tau+1)}}
,
\nonumber \\
E^{(\tau+1)}&=&\frac{1}{y_1^{(\tau+1)}}E^{(\tau)}+\frac{1}{(d-3)}\frac{1}{y_1^{(\tau+1)}}\left(x_1^{(\tau+1)}-\frac{1}{dx_1^{(\tau+1)}}\right)
\eea
with initial conditions 
$A^{(1)},B^{(1)},C^{(1)},D^{(1)},E^{(1)}$ which can be found by inserting $x_1^{(1)}=1-\mu$ and $y_1^{(1)}=[1+\frac{d+1}{d(1-\mu)}]$ in Eq. (\ref{defAdg3}).
The RG flow resulting from these equations can be studied numerically for any finite dimension $d$.
In particular we can find the relevant fixed point and the maximum eigenvalue $\lambda$ of the Jacobian of the RG equation at the relevant fixed point.
Finally by studying the density of eigenvalues $\rho(\mu)$ for $\mu\ll1$ we can derive the spectral dimension 
\bea
d_s=2\frac{\ln(d+1)}{\ln \lambda}.
\eea
The spectral dimension $d_s$ of $d$-dimensional pseudo-fractal networks with $2\leq d\leq 20$ is reported in Table \ref{ds:table}  together with the spectral dimension of the Apollonian networks of the same topological dimension $d$.
Finally  we observe that in the large $d$ limit the RG equations (\ref{RGhd}) for the pseudo-fractal network have the same leading term as the RG equations (\ref{RGhd3}) for the Apollonian networks. Therefore for $d\gg 1$, the largest eigenvalue $\lambda$ close to the non-trivial fixed point will have the same leading behavior. Indeed we can check numerically (see Fig. \ref{lambdadiff}) that the largest eigenvalue $\lambda$ satisfy
\bea
\lambda=\frac{d^2}{d^2-d-1}+{\mathcal O}(d^{-2}),
\eea
where $d^2/(d^2-d-1)$ is the leading eigenvalue of the RG flow for the Apollonian network.
It follows that for the pseudo-fractal network the  leading term for the spectral dimension is
\bea
d_S \simeq 2 d \ln (d+1).
\eea
This is confirmed by numerical results shown in Fig.~\ref{dsspeudo}.
\begin{table}
\caption{\label{ds:table}Numerical values for the spectral dimension $d_s$ of the $d$-dimensional Apollonian network and of the $d$-dimensional pseudo-fractal network.}
\begin{indented}
\item[]\begin{tabular}{@{}lll}
\br
$d$&$d_s$ Apollonian network& $d_s$ pseudo-fractal network\\
\mr
2&2&3.16993\\
3&3.73813& 5.31562\\
4&7.39962&8.3761\\
5& 11.729&12.7543\\
6&16.5732& 17.8447\\
7& 21.8337&23.421\\
8&27.4423&29.3665\\
9&33.3496&35.610\\
10&39.5179&42.104\\
11& 45.9179 &  48.8146\\
12&52.5262&55.7169\\
13&59.3233&62.7913\\
14&66.2933&70.0226\\
15&73.4226&77.3979\\
16&80.6995&84.9067\\
17&88.1141&92.5401\\
18&95.6574&100.29\\
19& 103.322& 108.150\\
20& 111.100&116.114\\
\br
\end{tabular}
\end{indented}
\end{table}


\subsection{Comparison to NGF with $s=0,1$}

\begin{figure}[htbp]
\begin{center}
\includegraphics[width=0.9\textwidth]{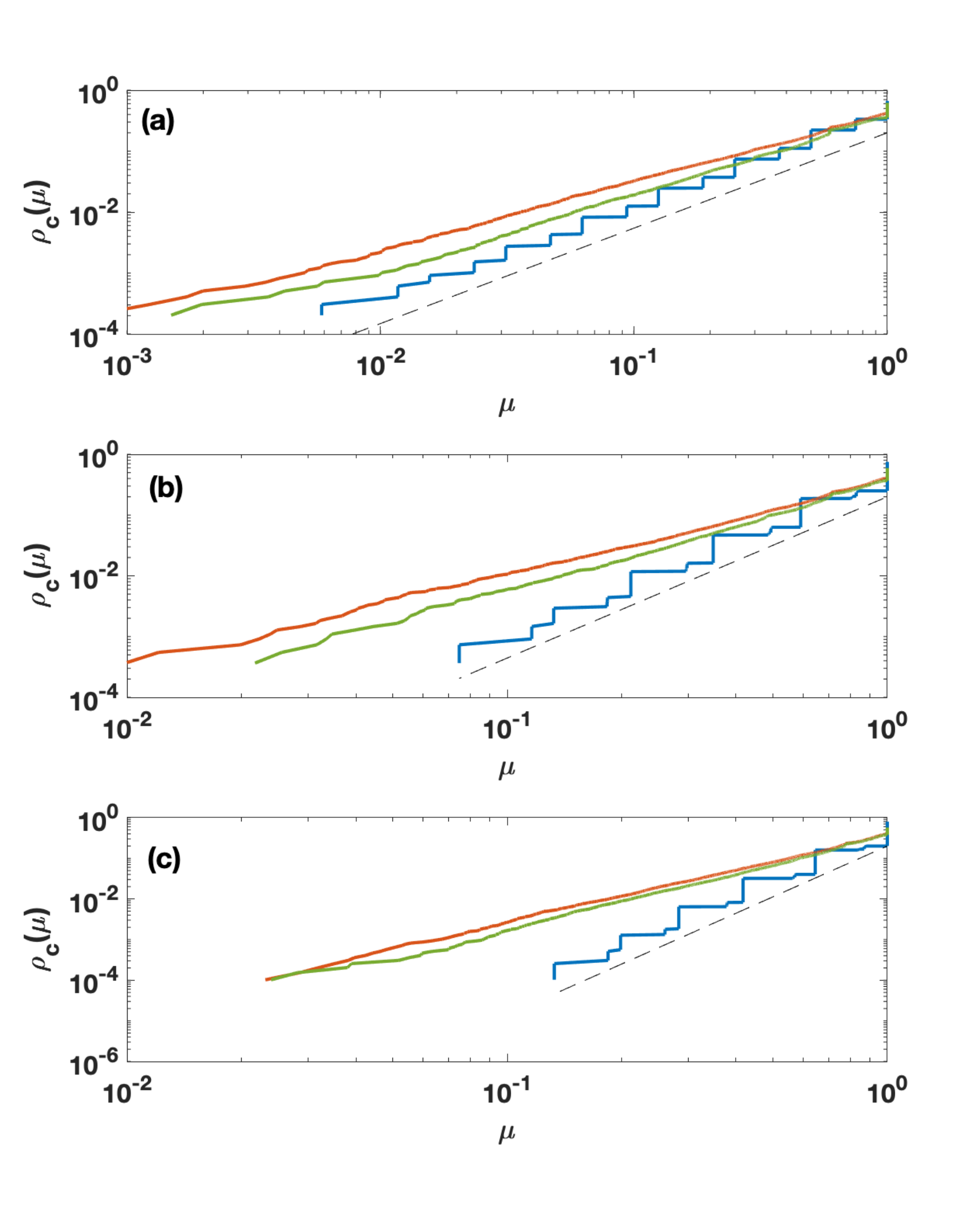}
\caption{Cumulative distribution of the eigenvalues $\rho_c(\mu)$ for the pseudo-fractal network in $d=2$(panel a), $d=3$ (panel b) and $d=4$ (panel c) (shown in blue) are  compared with the theoretical predictions of the spectral dimension (black dashed line) and with the cumulative distribution of the eigenvalues for the NGF with flavor $s=0$ (red lines) and $s=1$ (green lines).}
\label{pseudofractal_NGF}
\end{center}
\end{figure}

In this paragraph we compare the results obtained numerically for the pseudo-fractal networks with the theoretical predictions for $d=2,d=3$ and $d=4$. The numerical evaluation of the cumulative distribution $\rho_c(\mu)$ clearly show that the pseudo-fractal networks of topological dimension $d=2,3$ and $d=4$ display the predicted spectral dimension (see Fig.~\ref{pseudofractal_NGF}).
Moreover, in Fig.~\ref{pseudofractal_NGF} we also  compare the results  for the cumulative distribution of the eigenvalues $\rho_c(\mu)$ of the pseudo-fractal network to the one  obtained for NGFs with flavor $s=0$ and $s=1$. We found that the randomness present in the NGF significantly lowers the spectral dimension of the NGF. At the same time we observe that the degeneracies of eigenvalues present in the pseudo-fractal network are less pronounced for the NGFs.

\section{Conclusions}

 In this work we have studied the spectral dimension of the skeleton of simplicial complexes with distinct geometrical properties: the Apollonian networks, the pseudo-fractal networks and the Network Geometry with Flavor. The Apollonian networks are non-amenable hyperbolic manifolds, the pseudo-fractal networks are non-amenable branching networks, and the Network Geometry with Flavor (NGF) are non-amenable network structures that can be either be hyperbolic manifolds (for $s=-1$) or branching   (for $s=0,s=1$) simplicial complexes. However while Apollonian networks and pseudo-fractal networks are deterministic the NGF describes a stochastic model.
We used the RG approach to predict  the spectral dimension of Apollonian and pseudo-fractal networks. We have  obtained the functional dependence of the  spectral dimension $d_S$ on the topological dimension $d$ of their underlying simplicial complex structure and have found  that for large value of $d$, the spectral dimension $d_S$ scales like $d_S\simeq 2d\ln d$ for both the Apollonian and the pseudo-fractal networks. We have shown that the spectral dimension of the planar Apollonian network of dimension $d=3$ is $d_S=2\ln 3/\ln(9/5)$.
Finally we have studied the effect of randomness on the spectral properties of the networks by comparing the spectrum of the Apollonian network to the spectrum of the NGF with $s=-1$ and the spectrum of the speudo-fractal network to the spectrum of the NGF with $s=0$ and $s=1$. We have found numerically the  intuitively reasonable result that randomness can only reduce the spectral dimension of the underlying lattice where some nodes and links are removed.
We hope that this work will stimulate further interest in the relations between network geometry and spectral dimension.

Finally we observe that this work can be extended in different directions. On one side, an extension of the RG technique to address the spectral dimension of random topologies might be challenging but would be very much welcome, as real networks are typically driven by a stochastic evolution. On the other side it would be very interesting to investigate further the  spectrum of highly symmetrical network structures like the one considered in this paper and predict the degeneracies of eigenvalues.

\end{document}